\documentclass[12pt,preprint]{aastex}
\shorttitle{}
\shortauthors{Nesvorn\'y et al.}

\begin{document}
\baselineskip 19.pt

\title{Evidence for Very Early Migration of the Solar System Planets\\
from the Patroclus-Menoetius binary Jupiter Trojan}

\author{David Nesvorn\'y$^1$, David Vokrouhlick\'y$^{1,2}$, William F. Bottke$^1$,\\ 
Harold F. Levison$^1$}
\affil{(1) Department of Space Studies, Southwest Research Institute,\\
1050 Walnut St., Suite 300, Boulder, CO, 80302, USA}
\affil{(2) Institute of Astronomy, Charles University, V Hole\v{s}ovi\v{c}k\'ach 2, 
CZ--18000 Prague 8, Czech Republic}

{\bf
The orbital distribution of trans-Neptunian objects provides strong evidence for 
the radial migration of Neptune [1,2]. The outer planets' orbits are thought to have 
become unstable during the early stages [3] with Jupiter having scattering encounters 
with a Neptune-class planet [4]. As a consequence, Jupiter jumped inward by a fraction 
of an au, as required from inner solar system constraints [5,6], and obtained its 
current orbital eccentricity. The timing of these events is often linked to the lunar Late Heavy Bombardment 
that ended $\sim$700 Myr after the dispersal of the protosolar nebula ($t_0$) [7,8]. 
Here we show instead that planetary migration started shortly after $t_0$. Such early migration 
is inferred from the survival of the Patroclus-Menoetius binary Jupiter Trojan [9]. The binary 
formed at $t \lesssim t_0$ [10,11] within a massive planetesimal disk once located beyond 
Neptune [12,13]. The longer the binary stayed in the disk, the greater the likelihood that 
collisions would strip its components from one another. The simulations of its 
survival indicate that the disk had to have been dispersed by migrating planets 
within $\lesssim$100 Myr of $t_0$. This constraint implies that the planetary migration 
is unrelated to the formation of the youngest lunar basins.} 

Jupiter Trojans (JTs) are a population of small bodies with orbits near Jupiter [14]. 
They hug two equilibrium points of the three-body problem, known as $L_4$ and $L_5$, with 
semimajor axes $a \simeq 5.2$~au, eccentricities $e<0.15$, and inclinations $i<40^\circ$. 
Dynamical models suggest that JTs formed in the outer planetesimal disk between $\sim$20 au 
and 30 au and were implanted onto their present orbits after having a series of scattering 
encounters with the outer planets [12,13]. This resolves a long-standing conflict between the 
previous formation theories that implied $i<10^\circ$ and high orbital inclinations of JTs. 
The formation of JTs at 20-30 au is reinforced by their similarities to trans-Neptunian objects 
(TNOs; e.g., the absolute magnitude distribution and colors [15]).  

(617) Patroclus and Menoetius stand out among the 25 largest JTs with diameters $D>100$ km 
[16,17] as a curious pair of gravitationally bound bodies with binary separation 
$a_{\rm B} \simeq 670$ km. The formation of the Patroclus-Menoetius (P-M) binary is thought 
to be related to the accretion processes of small bodies themselves [10,11]. The formation model 
from ref. [10] implies that the P-M binary formed by capture in a dynamically cold disk at 
$t \sim t_0$. In [11], it formed at $t<t_0$. The P-M binary provides an interesting constraint on the 
early evolution of the solar system. Two conditions must be satisfied: (i) the P-M binary survived 
collisional grinding in its parent planetesimal disk at 20-30 au, which sets limits on 
the disk lifetime; (ii) it survived planetary encounters during its
transport from 20-30 au to 5.2 au, which sets limits on the nature of gravitational 
scattering events during encounters.  

We first evaluated the dynamical effect of planetary encounters [18] to demonstrate the 
plausibility of the implantation model. For that, we repeated numerical simulations from [13] 
(see Methods) and monitored all encounters between disk planetesimals and planets. The planetesimals 
that evolved onto JT orbits were selected for further use. Each selected body was then assumed to be a binary with 
the total mass $m_{\rm B}=1.2 \times 10^{21}$ g [19]. The initial eccentricities of binary orbits, 
$e_{\rm B}$, were set to zero and the inclinations were chosen at random (assuming the isotropic 
orientation of the orbit-normal vectors). The binary orbits were propagated through encounters. 
We varied the initial binary semimajor axis, $a_{\rm B}$, to determine how binary survival 
depends on the initial separation of binary components.

The binary survival is sensitive to $a_{\rm B}$ (Figure 1). Most tight, P-M--mass
binaries with $a_{\rm B}<1500$ km survive, while most wide binaries with $a_{\rm B}>1500$ km 
do not. The wide binaries become unbound during close planetary encounters. Specifically, 
when the planetocentric Hill radius of the binary, $r_{\rm Hill,B}=q(m_{\rm B}/3m_{\rm pl})^{1/3}$, 
where $q$ is the distance of the closest approach and $m_{\rm pl}$ is the planet mass, becomes 
smaller than the binary separation; i.e., $r_{\rm Hill,B} < a_{\rm B}$ [20]. For encounters with 
Jupiter, this condition works out to be $q<1680a_{\rm B}$ or $q<2.5\times10^6$ km for $a_{\rm B}=1500$ 
km, which is $\simeq$0.05 of Jupiter's Hill sphere. The removed binaries become unbound or 
collapse (typically because $e_{\rm B}$ becomes large). In 12-15\% of cases, the bodies form 
a contact binary. This process may explain (624) Hektor, which is thought to be a contact binary [21]. 
For reference, the contact binary fraction among JTs is estimated to be 13-23\% [22].
 
The survival probability of the P-M binary during planetary encounters is $\simeq$72\%.
Compared to other, nearly-equal-size binaries among TNOs [23], the P-M binary with 
$a_{\rm B}/(R_1+R_2)\simeq6.2$, where $R_1$ and $R_2$ are the radii of binary components, stands out 
as unusually compact (TNO binaries have $10 \lesssim a_{\rm B}/(R_1+R_2)<1000$). This trend is 
consistent with what we know, because the P-M binary in the TNO region would not be spatially resolved 
by telescopic observations and wide TNO binaries would not survive dynamical implantation 
onto a JT orbit (Figure 1). We predict that tight, P-M--class binaries will be found in the 
TNO region when observations reach the $\simeq$0.02 arcsec resolution needed to resolve them 
(the current limit with HST is $\simeq$0.06 arcsec [23]).

The outer planetesimal disk at 20-30 au, in which the P-M binary formed, is thought to have 
been massive (total estimated mass $M_{\rm disk}\simeq20$ $M_\oplus$, where $M_\oplus\simeq6\times10^{27}$ 
g is the Earth mass), as inferred from planetary migration/instability simulations
[4], slow migration of Neptune required to explain the inclination distribution of TNOs [24], and 
the capture probability of JTs [13]. The massive disk was subject to intense collisional grinding 
by impacts between planetesimals. The survival of the P-M binary in such a hostile environment is an 
important constraint on the disk lifetime, $t_{\rm disk}$, defined as the time interval between $t_0$ and 
the start of Neptune's migration. 

This factor can be illustrated in the following example. 
Assume that a small projectile, carrying the linear momentum $p=m_{\rm i} v_{\rm i}$, where 
$m_{\rm i}$ is the projectile mass and $v_{\rm i}$ is the impact speed, hits one of the components of the 
P-M binary. In the limit of a fully inelastic collision, the momentum $p$ is transferred and 
the binary orbit must change. The magnitude of this change, $\Delta a_{\rm B}$, is 
$\Delta a_{\rm B}/a_{\rm B} \sim (m_{\rm i}/m_{\rm B})(v_{\rm i}/v_{\rm B})$, where $v_{\rm B}$ is the 
orbital speed of the binary orbit. The P-M binary has $v_{\rm B} \simeq 11$ m s$^{-1}$. 
Thus, to have $\Delta a_{\rm B}/a_{\rm B} \gtrsim 1$, the impactor mass must exceed $m_{\rm i} \sim 
0.01 m_{\rm B}$, where we assumed $v_{\rm i} = 1$ km s$^{-1}$. The specific kinetic energy of such an 
impactor is $Q=m v_{\rm i}^2/2m_{\rm B} \simeq 10^8$ erg g$^{-1}$, which is $\sim$10 times lower than 
the specific energy for the catastrophic disruption ($Q^*_{\rm D} \sim 10^9$ erg g$^{-1}$ for a 
100-km-class ice target; [25]). We thus see that relatively small, sub-catastrophic impacts on 
the P-M binary can dislodge Patroclus and Menoetius from their mutual orbit.

To study this process, we used a previously developed collision code (see Methods). 
The collisional evolution of the outer planetesimal disk is excessive for long disk lifetimes. By 400 Myr, 
the disk mass is $<$10 $M_\oplus$ and the number of $D>10$ km planetesimals drops to $\sim 2\times10^8$ 
(Supplementary Figure 3). The former is inconsistent with the disk mass inferred from ref. [4], and 
the latter is more than an order of magnitude below the expectation based on the JT capture 
model [13]. These problems cannot be resolved by increasing the initial disk mass, because more massive 
disks grind faster and the survival of the P-M binary in a more massive disk would be problematic.
Here we adopted the strong ice disruption scaling laws from ref. [25]. Weaker versions of these laws, 
which may be more realistic for JTs/TNOs, would make the problems discussed here even worse.    

We found that P-M binary survival is sensitive to $t_{\rm disk}$ (Figure 2). For example, 
for $t_{\rm disk} = 400$ Myr and 700 Myr, which were the two cases suggested in 
the past to explain the lunar Late Heavy Bombardment (LHB) [7,8], the P-M survival probabilities 
are $7 \times 10^{-5}$ and $2 \times 10^{-7}$, respectively. Assuming a 100\% initial 
binary fraction, and adopting the 72\% dynamical survival probability computed previously, we 
find that having one P-M binary among the 25 largest JTs with $D>100$ km would be a $<$0.002 probability event 
if $t_{\rm disk}\geq400$ Myr. The long-lived disks can therefore be ruled out at the 99.8\% confidence 
level. In reality, the confidence is even greater because: (i) not all planetesimals formed as 
binaries, and (ii) binaries that formed with $a_{\rm B}>1000$ km cannot be the progenitors of the 
tight P-M binary (Supplementary Figure 1).

For $t_{\rm disk}<100$ Myr, the P-M survival probability against impacts is $>$10\%, indicating that  
short-lived planetesimal disks may be plausible. To demonstrate this, we adopted $t_{\rm disk}=0$ and considered 
the case when Neptune migrates into the planetesimal disk immediately after $t_0$. The impact probability 
and $v_{\rm i}$ were evaluated as a function of time from the $N$-body simulations of JT capture [13]. 
The changing conditions were implemented in our collisional code (see Methods), 
which was then used to determine the collisional survival of the P-M binary over the past 4.6 Gyr. 
We found that, to fit the present size distribution of JTs, the shape of the size distribution 
at $t_0+t_{\rm disk}$ must have been similar to the present one for $D>10$ km. The cumulative size 
distribution of JTs for $10<D<100$~km can be approximated by $N(>\!\!D)\propto D^{-\gamma}$ with $\gamma \simeq 2$. 
For $D<10$ km, the slope of JTs is shallower [26]. This is well reproduced in our simulations, 
where $D<10$ km JTs are removed by disruptive impacts (Figure 3). 

The survival probability of the P-M binary is found to be 87\% for $t_{\rm disk}=0$ (Figure~4). Coupled with 
the dynamical survival from Figure 1, the combined probability is 62\%. Thus, roughly two in three primordial 
binaries with the P-M mass and separation would have survived to the present time (for $t_{\rm disk}=0$). 
This result can be used to estimate the occurrence rate of the P-M binaries in the original planetesimal 
disk. Given that P-M is the only known binary system among 25 JTs with $D>100$ km, the primordial binary 
occurrence rate for $a_{\rm B}<1000$ km was at least 6.5\% ($t_{\rm disk}>0$ would imply larger initial fractions). 
These results constitute the first constraint on the formation of tight, equal-size binaries in the outer solar system. 
For comparison, about 30\% of dynamically cold TNOs are thought to be {\it wide} binaries ($a_{\rm B}>1000$ km; 
[23]). 

The results reported here have important implications for the early evolution of the solar system. They show
that giant planet migration cannot be delayed to $\sim$400-700 Myr after the dispersal of the protosolar 
nebula (99.8\% confidence). This undermines the relation between the late planetary migration/instability 
and LHB suggested in [7,8], and alleviates problems with the orbital excitation in   
the terrestrial planet region [5,27]. Instead, we find that the planetary migration/instability happened early, and 
the asteroid/comet projectiles bombarded the terrestrial worlds early as well. With $t_{\rm disk} \leq 100$~Myr, 
the asteroid projectiles are estimated to have produced only $<$1/10 of large lunar craters, and fell short 
by a factor of $\sim$100 to explain the formation of the Orientale/Imbrium basins at $\simeq$3.9 Ga [28]. 
These arguments give support to the possibility that most LHB impactors originated in the terrestrial planet 
region [29,30].

\noindent
{\bf Corresponding author}\\
David Nesvorn\'y\\
Southwest Research Institute\\
1050 Walnut St., Suite 300\\
Boulder, Colorado 80302\\
Phone: (303) 546-0023\\
Email: davidn@boulder.swri.edu     

\noindent
{\bf Acknowledgements}\\
This work was funded by NASA'a SSERVI and Emerging Worlds programs, and the Czech Science Foundation
(grant 18-06083S). We thank A. Morbidelli for helpful suggestions.  

\noindent
{\bf Author contributions}\\
D.N. had the original idea, performed the simulations, and prepared the manuscript for publication. 
D.V. developed the binary module in the collision code and the N-body code for planetary encounters. D.V., 
W.F.B. and H.F.L. suggested additional tests and helped to improve the manuscript.

\clearpage
\begin{figure}
\epsscale{0.9}
\plotone{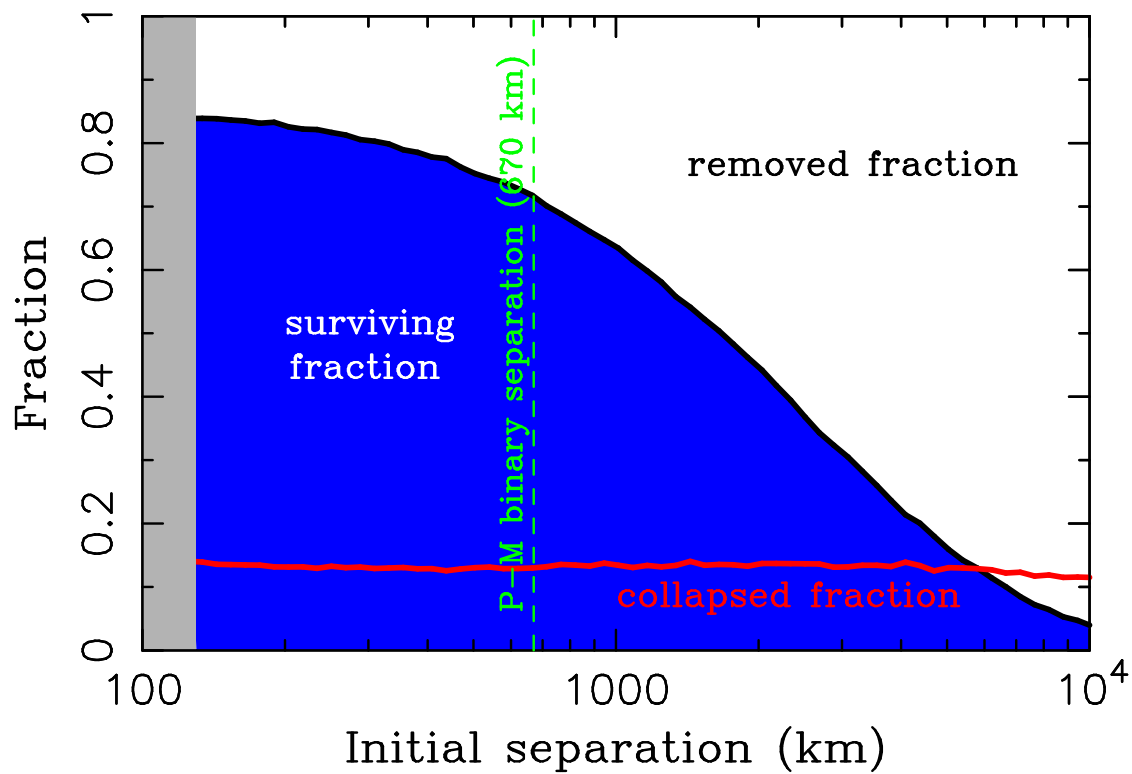}
\caption{The dynamical survival of binaries before their implantation onto JT orbits. The P-M 
binaries with $a_{\rm B}=670$ km (green line) survive in 72\% of cases, become unbound in 15\% of cases, and 
collapse into a contact binary in 13\% of cases (red line). The grey area displays the conditions 
for which the P-M components are in contact.}
\label{dyn}
\end{figure}

\clearpage
\begin{figure}
\epsscale{0.6}
\plotone{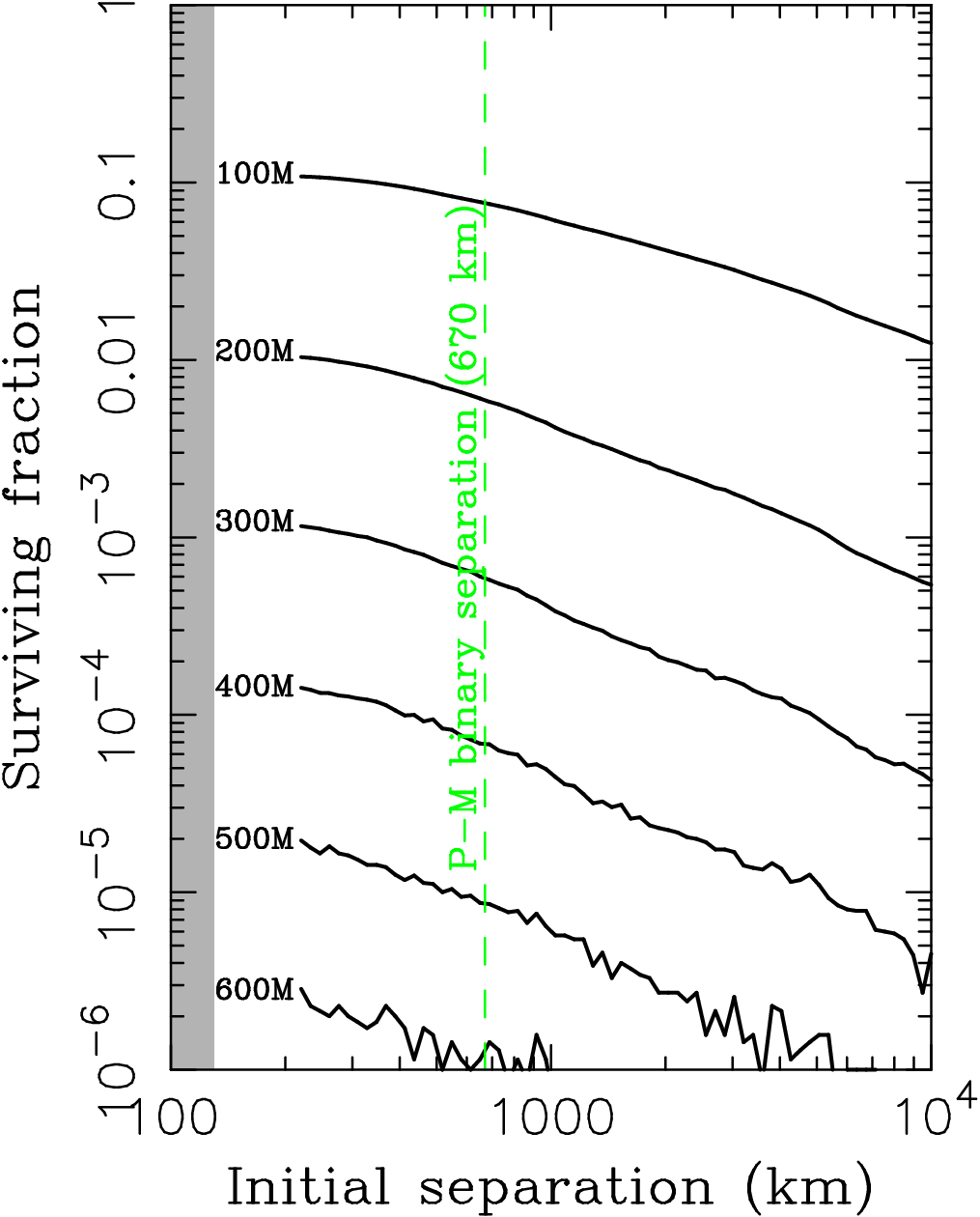}
\caption{The collisional survival of binaries in the outer planetesimal disk. The surviving fraction is 
shown for the P-M mass binaries as a function of the initial separation and disk lifetime (labels denote
$t_{\rm disk}$ in Myr; e.g., 100M corresponds to $t_{\rm disk}=100$ Myr). For the 
P-M binary separation and $t_{\rm disk} \geq 400$ Myr, the survival probability is $<10^{-4}$.
The grey area displays the conditions for which the P-M components are in contact.}
\label{col}
\end{figure}

\clearpage
\begin{figure}
\epsscale{0.6}
\plotone{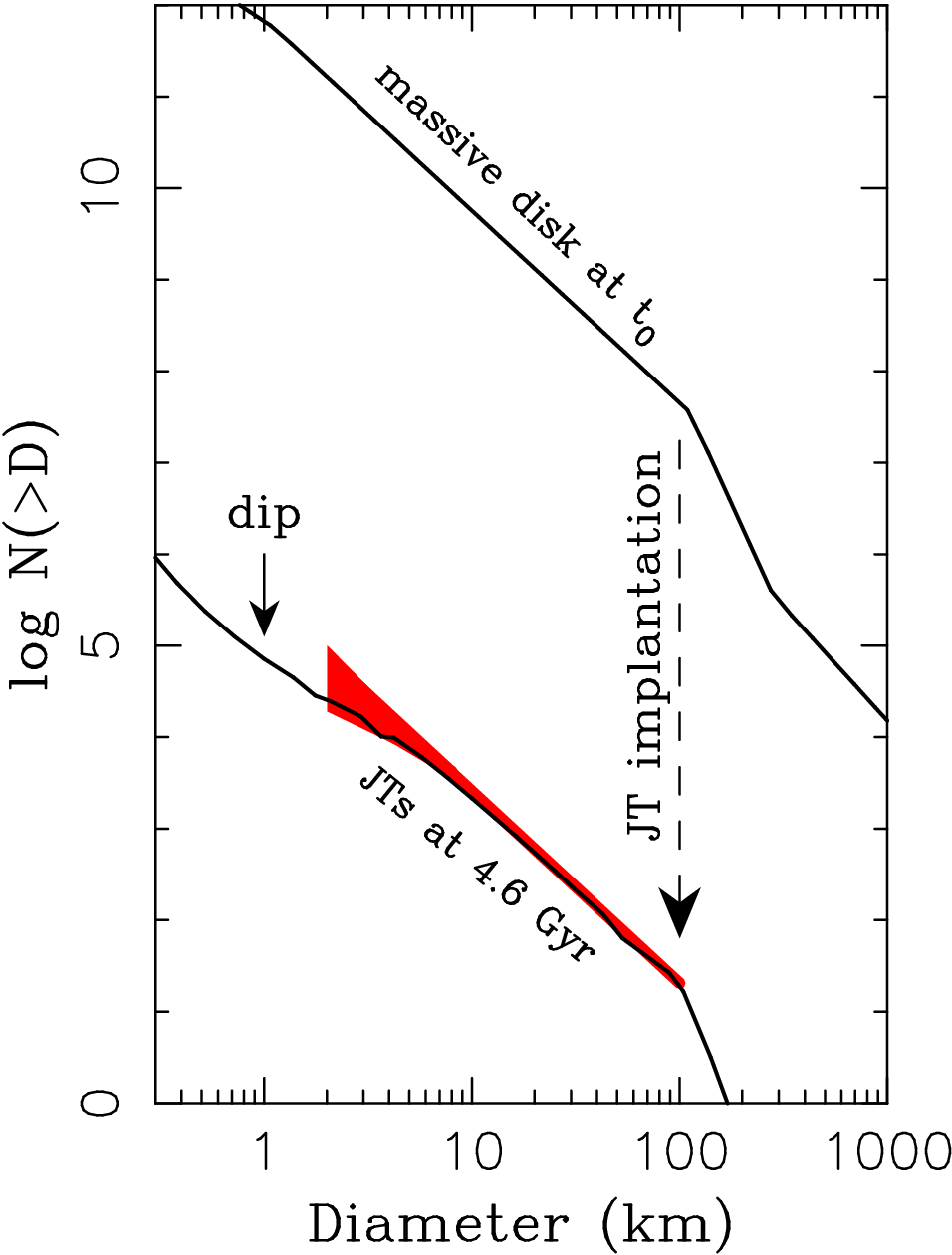}
\caption{The size distribution of JTs. Initially, a 20 $M_\oplus$ planetesimal disk was placed at 
20-30 au. During the disk dispersal, here assumed to have started at $t_0$, a small fraction of planetesimals 
($\simeq 5\times 10^{-7}$; [13]) was implanted onto JT orbits at 5.2 au. Here we used our collisional 
code to follow the collisional grinding of JTs at all stages of evolution. The final population of JTs is 
a scaled down version of the massive disk, except for $D<10$ km, where the collisional evolution produced 
a dip in the size distribution. This result is consistent with observations, here shown in red, which 
indicate a changing slope of JTs below $\sim$10 km [26].}
\label{sfd}
\end{figure}

\clearpage
\begin{figure}
\epsscale{0.9}
\plotone{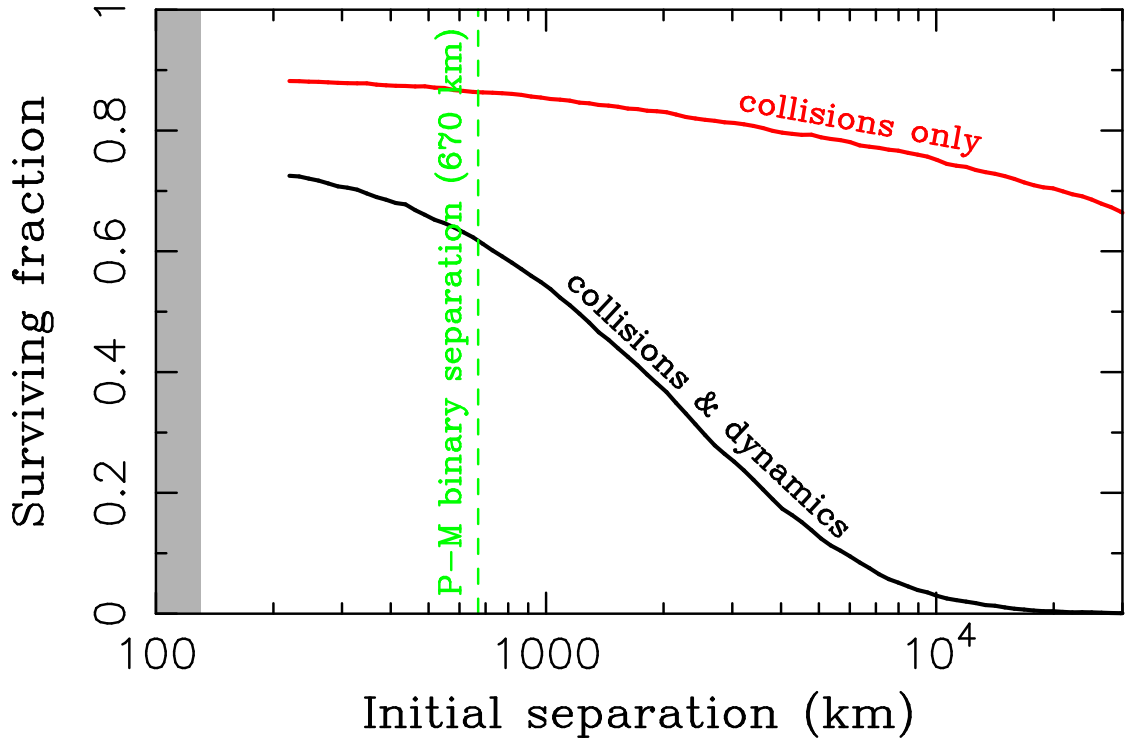}
\caption{The survival of binaries in the case when the planet migration was initiated 
immediately after $t_0$ (i.e., $t_{\rm disk}=0$). The red line shows the collisional survival 
of P-M mass binaries as a function of $a_{\rm B}$. The black line combines the
collisional survival with the dynamical survival from Figure 1. It expresses our expectation
for the fraction of the P-M--class binaries that should have survived to the present time
($t_{\rm disk}>0$ would imply lower fractions; Figure 2). The grey area displays the conditions 
for which the P-M components are in contact.}
\label{final}
\end{figure}

\clearpage

\noindent{\bf Methods}\\
\noindent{\bf Dynamical Effects of Planetary Encounters on Binaries}\\
We make use of the previously published simulations of JT capture [13] to evaluate the dynamical effect of 
planetary encounters on the P-M binary. To study capture, ref. [13] adopted three simulations of planetary 
instability/migration [4]. A shared property of the selected runs is that Jupiter undergoes a series of 
planetary encounters with an ice giant. The orbit of Jupiter evolves in discrete steps as a result of 
these encounters (the so-called {\it jumping-Jupiter model}). JTs are captured in the jumping-Jupiter 
model when Jupiter's Lagrange points become radially displaced by scattering events and fall into a 
region populated by planetesimals. The captured population was shown to provide a good match to both 
the orbital distribution of JTs and their total mass.

In [13], planetesimals were initially distributed in an outer disk extending from just beyond the initial 
orbit of Neptune at 22 au to 30~au. The outer extension of the disk beyond 30~au was ignored because 
various constraints indicate that a large majority of planetesimals started at $<$30 au (e.g., [31]). 
Also, the JT capture probability from the $>$30 au region is exceedingly small. The simulations were 
performed with a modified version of the symplectic $N$-body integrator known as {\it Swift} [32]. 
All encounters of planetesimals to planets were recorded. This was done by monitoring the distance of 
each planetesimal from Jupiter, Saturn, Uranus and Neptune, and recording every instance when the 
distance dropped below 0.5 $R_{{\rm Hill},j}$, where $R_{{\rm Hill},j}$ are the Hill radii of planets ($j=5$ 
to 8 from Jupiter to Neptune). We made sure that more distant encounters do not have any significant 
effect on the P-M binary. This was done by verifying that the results do not change when more 
distant encounters are accounted for.  

The sizes of P-M binary components were obtained from the occultation observations in [17]: $127 
\times 117 \times 98$ km for Patroclus and $117 \times 108 \times 90$ km for Menoetius. 
A volume-equivalent spherical size corresponds to diameters $D_1=113$ km for Patroclus and 
$D_2=104$ km for Menoetius. These dimensions and the total mass $1.2\times10^{21}$ g from [19] 
imply the system density $\simeq$0.88 g cm$^{-3}$. These are the values adopted in the main text. 
To study the dependence of our results on binary separation, the initial binary semimajor axis, 
$a_{\rm B}$, was treated as a free parameter ($200<a_{\rm B}<10^5$ km; for reference, the P-M 
binary has $a_{\rm B} \simeq 670$ km). The initial orbits were assumed to be circular 
(eccentricity $e_{\rm B}=0$) and randomly oriented in space.  

We used the model from [33] to compute the effect of planetary encounters on binaries. Each binary 
planetesimal was traced through recorded encounters using the Bulirsch-Stoer integrator that we 
adapted from Numerical Recipes [34]. The Sun and other planets not having an encounter 
were neglected. First, we integrated the center of mass of a binary planetesimal backward from the 
closest approach until the planetocentric distance reached 3 $R_{{\rm Hill},j}$. Second, we replaced it 
by the actual binary and integrated forward through the encounter. The second integration was 
stopped when the binary reached 3 $R_{{\rm Hill},j}$. The final binary orbit was used as the initial 
orbit for the next encounter. The algorithm was iterated over all recorded encounters. 

Collisions between binary components were monitored. If a collision was detected, the integration was 
stopped and the code reported the impact speed and angle. Hyperbolic binary orbits were deemed to be 
unbound. The final values of $a_{\rm B}$ and $e_{\rm B}$ were recorded for the surviving 
binaries. We found that, in the regime corresponding to the P-M binary separation ($a_{\rm B}<1000$~km), 
the final separation is generally a good proxy for the initial separation. For example, in all cases 
that ended with $a_{\rm B}=670$ km, only 1\% of the binary orbits started with $a_{\rm B}<380$ km or 
$a_{\rm B}>970$ km (Supplementary Figure 1). This justifies our assumption that the P-M binary started 
with $a_{\rm B} \sim 670$ km.  
 
\noindent{\bf Collisional Evolution}\\
The mutual orbit of a binary can be affected by small impacts into its components [35]. To study this 
process, we used the code known as {\it Boulder} [36,37]. The {\it Boulder} code employs a statistical 
algorithm to track the collisional fragmentation of planetesimal populations. Here we briefly 
highlight the main points and differences with respect to refs. [36] and [37]. 

For each collision, the code computes the specific impact energy $Q$ and the critical impact energy 
$Q^*_{\rm D}$ (see ref. [25] for definitions]. Based on the value of $Q/Q^*_{\rm D}$ and available scaling 
laws (e.g., [25]), it then determines the masses of the largest remnant and largest fragment, and the 
power-law size distribution of smaller fragments (e.g., [38]). The $Q^*_{\rm D}$ function in {\it Boulder} 
was set to be intermediate between impact simulations with strong [25] and weak ice [39]. To achieve
this, we multiplied $Q^*_{\rm D}$ from [25] by a factor $f_Q$, where $f_Q=1$, 0.3 and 0.1 was used in 
different tests. The impact experiments with highly porous targets suggest that the $Q^*_{\rm D}$ values 
can be slightly higher than those found for strong ice [40]. This result reflects the dissipative 
properties of material porosity. We verified that using scaling laws from [40] in the {\it Boulder} code 
gives results that are very similar to those obtained with $f_Q=3$. We therefore tested $f_Q=3$ as well.

The main input parameters are: the (i) initial size distribution of simulated populations, 
(ii) intrinsic collision probability $P_i$, and (iii) mean impact speed $v_i$. The initial size distribution 
can be informed from JTs, which are observationally well characterized down to at least 5 km [26].
For $5\lesssim D \lesssim 100$ km, the cumulative size distribution $N(>\!\!D)$ is a power law 
$N(>\!\!D) \propto D^{-\gamma}$ with $\gamma \simeq 2$. Above $D \simeq 100$ km, the 
JT size distribution bends to a much steeper slope ($\gamma \sim 6$). There are 25 JTs with $D>100$ km 
[16]. For $D < 5$ km, the JT size distribution bends to a shallower slope with $\gamma < 2$ [26].
As we discuss in the main text (Figure 3), the shallow slope at small sizes suggests that JTs evolved 
through a stage of modest collisional grinding. 

The JT capture efficiency from the original planetesimal disk is well defined. Ref. [13] and our additional 
simulations suggest $P_{\rm capture} = (5 \pm 2) \times 10^{-7}$ (this is a probability that an outer disk 
planetesimal ends up on a stable JT orbit), where the error bars give the full range of values obtained 
in different simulations. We adopt $P_{\rm capture} = 5 \times 10^{-7}$ in this work. To construct the 
size distribution of planetesimals, the JT size distribution is divided by $P_{\rm capture}$ (Supplementary 
Figure 2). This gives $\simeq 6 \times 10^9$ planetesimals with $D>10$ km. The total mass of the 
reconstructed population is 20 $M_\oplus$, in agreement with [4]. 

As for $P_i$ and $v_i$, we performed two different tests. The first test was intended to replicate the 
collisional grinding of the outer planetesimal disk. In this case, we assumed that migrating Neptune 
removed the disk at $t_{\rm disk}$ after the dispersal of the protosolar nebula ($t_{0}$), and let the disk 
collisionally evolve over $t_{\rm disk}$. The dynamical state of the disk was taken from [41]. For example, 
at 300 Myr after $t_0$, the disk at 20-30~au is characterized by $P_i \simeq 8 \times 10^{-21}$ km$^{-2}$ 
yr$^{-1}$ and $v_i \simeq 0.4$ km s$^{-1}$ [42].

Collisional grinding of the outer planetesimal disk proceeds fast (Supplementary Figure 3). For 
$t_{\rm disk}>100$ Myr, the number of $D > 10$ km bodies is reduced at least tenfold and the total mass 
drops to $<$10 $M_\oplus$. These results are in conflict with the current size distribution of JTs, 
the planetesimal disk mass inferred in [4], and other constraints. The problem could potentially be 
resolved if we adopted a larger initial mass. We tested several possibilities along these lines. 
For example, we scaled up the reference size distribution by an additional factor to increase the 
initial mass to $>$20 $M_\oplus$. These tests failed because more massive disks grind faster and end up 
with $<$10 $M_\oplus$ for $t_{\rm disk}>100$ Myr. In other tests, we used a steeper slope for $D<100$ km in 
an attempt to obtain $\gamma \simeq 2$ as a result of collisional grinding. These tests failed as well 
for reasons similar to those described above. 

Using $f_Q>1$ does not resolve the problems discussed above. This is mainly due to two reasons.
First, very large values of $f_Q$ ($f_Q > 3$) are needed to significantly limit the effect of collisional 
grinding, but these values are probably too high to be realistic. Second, even if we use $f_Q > 3$, 
the number of $D \simeq 10$ km bodies is still reduced by a factor of $\sim$10. This is because, for 
the low impact speeds adopted here, the focusing factors can be large and small planetesimals are 
lost by efficiently accreting on the largest disk bodies. Given these unresolved issues, we decided to 
adopt the following scheme for our nominal simulation of impacts on the P-M binary. We used the 
reference size distribution (20 $M_\oplus$ initially) and switched off the fragmentation of 
planetesimals ($f_Q \gg 1$) and their accretion onto large bodies. In this case, the size distribution 
stayed approximately the same over the whole length of the simulation. This is arguably a very 
conservative assumption. Other schemes would require that the initial population was larger and decayed 
over time, implying more impacts overall. 

We tested many additional initial size distributions, including $\gamma \simeq 2$ for $D^*<D<100$ km and 
$\gamma < 2$ for $D<D^*$, where the transition diameter $D^*<100$~km was taken as a free parameter. 
This was done to verify whether the initial paucity of small projectiles would reduce the long-term exposure 
of the P-M binary to orbit-changing impacts. The end-member case of these models is the one with 
no $D<D^*$ bodies whatsoever (perhaps because they did not form). If fragmentation is switched off 
in this case ($f_Q \gg 1$), the size distribution remains unchanged and fails to match the present 
size distribution of JTs for $D<D^*$. If the fragmentation is switched on ($f_Q \sim 1$), the collisional 
cascade acts very quickly, within $\simeq$10 Myr, to produce a fragment tail with $\gamma \simeq 2$ below 
$D<D^*$ km (Supplementary Figure 4). The survival probability of the P-M binary is nearly the same 
in this case as in our nominal case, where the initial size distribution was extended to $D<D^*$ km 
with $\gamma \simeq 2$.

The second set of simulations with {\it Boulder} was done under the assumption that the outer planetesimal 
disk was dispersed by Neptune immediately after $t_0$ (i.e., $t_{\rm disk}=0$). The disk was assumed 
to have started dynamically cold ($e \simeq 0$ and $i \simeq 0$) or hot (Rayleigh distributions in $e$ 
and $i$). It was gradually excited after $t_0$, on a timescale of 10-30 Myr, by migrating Neptune. 
The \"Opik algorithm [43,44] and the simulations reported in [13] were used to compute $P_i$ and 
$v_i$ as a function of time (Supplementary Figure 5). We selected planetesimals that became captured as JT 
and monitored their collision probabilities and impact velocities with all other planetesimals. 
The $P_i$ and $v_i$ values were computed each $\delta t$ by averaging over the selected planetesimals,  
where $\delta t = 1$ Myr during the initial stages, when $P_i$ and $v_i$ change quickly, and 
$\delta t=10$-100 Myr later on. After approximately 200 Myr past $t_0$, the collision evolution of JTs 
is dominated by impacts {\it among} JTs. After this transition, $P_i = 7 \times 10^{-18}$ km$^{-2}$ 
yr$^{-1}$ and $v_i=4.6$ km s$^{-1}$ [45].  

\noindent{\bf Impacts on the P-M binary}\\
The binary module in {\it Boulder} [37] accounts for small, non-disruptive impacts on binary components, and 
computes the binary orbit change depending on the linear momentum of impactors. For each impact, 
the change of orbital speed, $\mathbf{v}_{\rm B}=\mathbf{v}_2-\mathbf{v}_1$, where $\mathbf{v}_1$ and $\mathbf{v}_2$ 
are the velocity vectors of components, is computed from the conservation of the linear momentum. This gives 
\begin{equation}
\delta \mathbf{v}_{\rm B} = {m_i \over m_2+m_i}\left( {1 \over 2} \mathbf{v}_i - {m_1 \over m_{\rm B}} \mathbf{v}_{\rm B}\right) 
\end{equation} 
for an impact on the secondary, and
\begin{equation}
\delta \mathbf{v}_{\rm B} = - {m_i \over m_1+m_i}\left( {1 \over 2} \mathbf{v}_i + {m_2 \over m_{\rm B}} \mathbf{v}_{\rm B}\right) 
\end{equation} 
for an impact on the primary, where $m_1$ and $m_2$ are the primary and secondary masses, $m_{\rm B}=m_1+m_2$, and
$m_i$ and $\mathbf{v}_i$ are the impactor's mass and velocity. 

The first term in Eqs. (1) and (2) corresponds to the transfer of the linear momentum. The factor 1/2 stands for the 
contribution of the impactor's linear momentum to the translational motion (as averaged over all impact geometries).
The rest of the linear momentum is consumed by the spin vector change of the impacted binary component. Note 
that this assumes that all collisions are completely inelastic. A larger yield would occur if it is established 
that the escaping ejecta affect the linear momentum budget [46], but we do not consider this effect here.  

The impact velocity vectors were assumed to be randomly oriented in the reference frames of binaries. We also 
factored in that impacts can happen at any orbital phase and averaged the binary orbit changes over the orientation 
and phase. The changes of orbital elements, $\delta a_{\rm B}$ and $\delta e_{\rm B}$, were computed from 
\begin{equation}
 {\delta a_{\rm B} \over a_{\rm B}} = \pm {1 \over \sqrt{3}} \, \frac{m_i v_i}{m_{\rm B} v_{\rm B}} \;  
\label{a}
\end{equation}
and 
\begin{equation}
 \delta e_{\rm B}  = \pm {1 \over 2} \sqrt{\frac{5}{6}} \, \eta \, \frac{m_iv_i}{m_{\rm B} v_{\rm B}} \; ,
\label{e} 
\end{equation}
where $v_i$ and $v_{\rm B}$ are the moduli of $\mathbf{v}_i$ and $\mathbf{v}_{\rm B}$, and $\eta^2=1-e_{\rm B}^2$. 
The $\pm$ sign in front of the right-hand sides indicates that the individual changes can be positive or 
negative. Equations (3) and (4) were implemented in the {\it Boulder} code. A similar expression can be 
obtained for inclinations [36], but we do not follow the inclination changes here.

\noindent{\bf Code availability}\\
The $N$-body integrator that was used in this work to record planetary encounters is 
available from https://www.boulder.swri.edu/\~{}hal/swift.html. The code was trivially 
modified to monitor the physical distance between test particles and planets, and 
record the planetocentric path of each particle during encounters. The $N$-body code 
that we used to track changes of the binary orbits is available from\\ 
http://www.boulder.swri.edu/\~{}davidn/Codes/. The {\it Boulder} code with the binary 
module was developed with internal SwRI funding and is proprietary.

\noindent{\bf Data availability}\\
The data that support the plots within this paper and other findings of this study are available 
from the corresponding author upon reasonable request.


\begin{thebibliography}

\bibitem[Hahn and Malhotra(2005)]{2005AJ....130.2392H} [1] Hahn, J.~M., Malhotra, R.\ Neptune's migration 
into a stirred-up Kuiper belt: a detailed comparison of simulations to observations.\ 
{\it Astron. J.} {\bf 130}, 2392-2414 (2005).

\bibitem[Levison et al.(2008)]{2008Icar..196..258L} [2] Levison, H.~F., Morbidelli, A., Van Laerhoven, C., 
Gomes, R., Tsiganis, K.\ Origin of the structure of the Kuiper belt during a dynamical instability in 
the orbits of Uranus and Neptune.\ {\it Icarus} {\bf 196}, 258-273 (2008).

\bibitem[Tsiganis et al.(2005)]{2005Natur.435..459T} [3] Tsiganis, K., Gomes, R., Morbidelli, A., Levison, H.~F.\ 
Origin of the orbital architecture of the giant planets of the Solar System.\ {\it Nature} {\bf 435}, 
459-461 (2005).

\bibitem[Nesvorn{\'y} and Morbidelli(2012)]{2012AJ....144..117N} [4] Nesvorn{\'y}, D., Morbidelli, A.\  
Statistical study of the early Solar System's instability with four, five, and six giant planets.\ 
{\it Astron. J.} {\bf 144}, 117 (2012).

\bibitem[Agnor and Lin(2012)]{2012ApJ...745..143A} [5] Agnor, C.~B., Lin, D.~N.~C.\ On the migration of Jupiter
 and Saturn: constraints from linear models of secular resonant coupling with the terrestrial planets.\  
{\it Astrophys. J.} {\bf 745}, 143 (2012).

\bibitem[Morbidelli et al.(2010)]{2010AJ....140.1391M} [6] Morbidelli, A., Brasser, R., Gomes, R., Levison, H.~F., 
Tsiganis, K.\ Evidence from the asteroid belt for a violent past evolution of Jupiter's orbit.\  
{\it Astron. J.} {\bf 140}, 1391-1401 (2010).

\bibitem[Gomes et al.(2005)]{2005Natur.435..466G} [7] Gomes, R., Levison, H.~F., Tsiganis, K., Morbidelli, A. 
Origin of the cataclysmic Late Heavy Bombardment period of the terrestrial planets.\ 
{\it Nature} {\bf 435}, 466-469 (2005). 

\bibitem[Bottke et al.(2012)]{2012Natur.485...78B} [8] Bottke, W.~F. {\it  et al.} An Archaean heavy bombardment 
from a destabilized extension of the asteroid belt.\ {\it Nature} {\bf 485}, 78-81 (2012).

\bibitem[Merline et al.(2001)]{2001IAUC.7741....2M} [9] Merline, W.~J. et al. S/2001 (617) 1.\ 
{\it International Astronomical Union Circular} {\bf 7741}, 2 (2001).

\bibitem[Goldreich et al.(2002)]{2002Natur.420..643G} [10] Goldreich, P., Lithwick, Y., Sari, R.\ 
Formation of Kuiper-belt binaries by dynamical friction and three-body encounters.\ {\it Nature} {\bf 420}, 
643-646 (2002). 

\bibitem[Nesvorn{\'y} et al.(2010)]{2010AJ....140..785N} [11] Nesvorn{\'y}, D., Youdin, A.~N., Richardson, D.~C.\ 
Formation of Kuiper belt binaries by gravitational collapse.\ {\it Astron. J.} {\bf 140}, 785-793 (2010).

\bibitem[Morbidelli et al.(2005)]{2005Natur.435..462M} [12] Morbidelli, A., Levison, H.~F., Tsiganis, K., Gomes, 
R.\ Chaotic capture of Jupiter's Trojan asteroids in the early Solar System.\ {\it Nature} {\bf 435}, 
462-465 (2005). 

\bibitem[Nesvorn{\'y} et al.(2013)]{2013ApJ...768...45N} [13] Nesvorn{\'y}, D., Vokrouhlick{\'y}, D., Morbidelli, 
A.\ Capture of Trojans by Jumping Jupiter.\ {\it Astrophys. J.} {\bf 768}, 45 (2013).


\bibitem[Emery et al.(2015)]{2015aste.book..203E} [14] Emery, J.~P., Marzari, F., Morbidelli, A., French, L.~M., 
Grav, T.\ The complex history of Trojan asteroids.\ {\it Asteroids IV}, 203-220 (2015).

\bibitem[Fraser et al.(2014)]{2014ApJ...782..100F} [15] Fraser, W.~C., Brown, M.~E., Morbidelli, A., Parker, A., Batygin, 
K.\ The absolute magnitude distribution of Kuiper belt objects.\ {\it Astrophys. J.} {\bf 782}, 100 (2014).

\bibitem[Grav et al.(2011)]{2011ApJ...742...40G} [16] Grav, T. {\it et al.} WISE/NEOWISE observations 
of the Jovian Trojans: preliminary results.\ {\it Astrophys. J.} {\bf 742}, 40 (2011). 

\bibitem[Buie et al.(2015)]{2015AJ....149..113B} [17] Buie, M.~W. {\it et al.} Size and shape from 
stellar occultation observations of the double Jupiter Trojan Patroclus and Menoetius.\ 
{\it Astron. J.} {\bf 149}, 113 (2015).

\bibitem[Parker and Kavelaars(2010)]{2010ApJ...722L.204P} [18] Parker, A.~H., Kavelaars, J.~J.\ 
Destruction of Binary Minor Planets During Neptune Scattering.\ {\it Astrophys. J.} {\bf 722}, L204-L208 (2010). 

\bibitem[Mueller et al.(2010)]{2010Icar..205..505M} [19] Mueller, M. {\it et al.} Eclipsing binary Trojan asteroid 
Patroclus: thermal inertia from Spitzer observations.\ {\it Icarus} {\bf 205}, 505-515 (2010).

\bibitem[Agnor and Hamilton(2006)]{2006Natur.441..192A} [20] Agnor, C.~B., Hamilton, D.~P.\ Neptune's capture of 
its moon Triton in a binary-planet gravitational encounter.\ {\it Nature} {\bf 441}, 192-194 (2006).


\bibitem[Marchis et al.(2014)]{2014ApJ...783L..37M} [21] Marchis, F. {\it et al.} The puzzling mutual 
orbit of the binary Trojan asteroid (624) Hektor.\ {\it Astrophys. J.} {\bf 783}, L37 (2014).

\bibitem[Sonnett et al.(2015)]{2015ApJ...799..191S} [22] Sonnett, S., Mainzer, A., Grav, T., Masiero, J., 
Bauer, J.\ Binary candidates in the Jovian Trojan and Hilda populations from NEOWISE light curves.\ 
{\it Astrophys. J.} {\bf 799}, 191 (2015).

\bibitem[Noll et al.(2008)]{2008ssbn.book..345N} [23] Noll, K.~S., Grundy, W.~M., Chiang, E.~I., Margot, J.-L., 
Kern, S.~D.\ Binaries in the Kuiper belt.\ {\it The Solar System Beyond Neptune}, 345-363 (2008).


\bibitem[Nesvorn{\'y}(2015)]{2015AJ....150...73N} [24] Nesvorn{\'y}, D.\ Evidence for slow migration of 
Neptune from the inclination distribution of Kuiper belt objects.\ {\it Astron. J.} {\bf 150}, 73 (2015).

\bibitem[Benz and Asphaug(1999)]{1999Icar..142....5B} [25] Benz, W., Asphaug, E.\ Catastrophic disruptions 
revisited.\ {\it Icarus} {\bf 142}, 5-20 (1999).

\bibitem[Wong and Brown(2015)]{2015AJ....150..174W} [26] Wong, I., Brown, M.~E.\ The color-magnitude distribution 
of small Jupiter Trojans.\ {\it Astron. J.} {\bf 150}, 174 (2015).


\bibitem[Kaib and Chambers(2016)]{2016MNRAS.455.3561K} [27] Kaib, N.~A., Chambers, J.~E.\ The fragility of the 
terrestrial planets during a giant-planet instability.\ {\it Mon. Not. R. Astron. Soc.} {\bf 455}, 3561-3569 
(2016).

\bibitem[Nesvorn{\'y} et al.(2017)]{2017AJ....153..103N} [28] Nesvorn{\'y}, D., Roig, F., Bottke, W.~F.\  
Modeling the historical flux of planetary impactors.\ {\it Astron. J.} {\bf 153}, 103 (2017).

\bibitem[Bottke et al.(2007)]{2007Icar..190..203B} [29] Bottke, W.~F., Levison, H.~F., Nesvorn{\'y}, D., Dones, L.\ 
Can planetesimals left over from terrestrial planet formation produce the lunar Late Heavy Bombardment?\ 
{\it Icarus} {\bf 190}, 203-223 (2007).

\bibitem[Morbidelli et al.(2018)]{2018Icar..305..262M} [30] Morbidelli, A. {\it et al.} 
The timeline of the lunar bombardment: Revisited.\ {\it Icarus} {\bf 305}, 262-276 (2018).

\end{thebibliography}

\begin{thebibliography}{}

\bibitem[Gomes et al.(2004)]{2004Icar..170..492G} [31] Gomes, R.~S., Morbidelli, A., 
Levison, H.~F.\ Planetary migration in a planetesimal disk: why did Neptune stop at 
30 AU? {\it Icarus} {\bf 170}, 492-507 (2004). 

\bibitem[Levison and Duncan(1994)]{1994Icar..108...18L} [32] Levison, H.~F., Duncan, M.~J.\ 
The long-term dynamical behavior of short-period comets.\ {\it Icarus} {\bf 108}, 18-36 (1994).

\bibitem[Nesvorn{\'y} et al.(2018)]{2018AJ....155..246N} [33] Nesvorn{\'y}, D., 
Parker, J., Vokrouhlick{\'y}, D.\ Bi-lobed Shape of Comet 67P from a Collapsed Binary.\ 
{\it Astron. J.} {\bf 155}, 246 (2018). 

\bibitem[Press et al.(1992)]{1992nrfa.book.....P} [34] Press, W.~H., Teukolsky, S.~A., Vetterling, 
W.~T., Flannery, B.~P.\ Numerical recipes in FORTRAN. The art of scientific computing.\ 
Cambridge: University Press (1992).

\bibitem[Petit and Mousis(2004)]{2004Icar..168..409P} [35] Petit, J.-M., Mousis, O.\ KBO binaries: 
how numerous were they? {\it Icarus} {\bf 168}, 409-419 (2004).

\bibitem[Morbidelli et al.(2009)]{2009Icar..204..558M} [36] Morbidelli, A., Bottke, W.~F., 
Nesvorn{\'y}, D., Levison, H.~F.\ Asteroids were born big.\ {\it Icarus} {\bf 204}, 
558-573 (2009). 

\bibitem[Nesvorn{\'y} et al.(2011)]{2011AJ....141..159N} [37] Nesvorn{\'y}, D., Vokrouhlick{\'y}, D., 
Bottke, W.~F., Noll, K., Levison, H.~F.\ Observed binary fraction sets limits on the extent 
of collisional grinding in the Kuiper belt.\ {\it Astron. J.} {\bf 141}, 159 (2011).

\bibitem[Durda et al.(2007)]{2007Icar..186..498D} [38] Durda, D.~D. {\it et al.} 
Size-frequency distributions of fragments from SPH/N-body simulations of asteroid impacts: 
Comparison with observed asteroid families.\ {\it Icarus} {\bf 186}, 498-516 (2007). 

\bibitem[Leinhardt and Stewart(2009)]{2009Icar..199..542L} [39] Leinhardt, Z.~M., Stewart, S.~T.\ 
Full numerical simulations of catastrophic small body collisions.\ {\it Icarus} {\bf 199},
542-559 (2009).

\bibitem[Jutzi et al.(2010)]{2010Icar..207...54J} [40] Jutzi, M., Michel, P., Benz, W., 
Richardson, D.~C.\ Fragment properties at the catastrophic disruption threshold: The effect 
of the parent body's internal structure.\ {\it Icarus} {\bf 207}, 54-65 (2010).

\bibitem[Levison et al.(2011)]{2011AJ....142..152L} [41] Levison, H.~F., Morbidelli, A., 
Tsiganis, K., Nesvorn{\'y}, D., Gomes, R.\ Late orbital instabilities in the outer planets 
induced by interaction with a self-gravitating planetesimal Disk.\ {\it Astron. J.} 
{\bf 142}, 152 (2011). 

\bibitem[Morbidelli and Rickman(2015)]{2015A&A...583A..43M} [42] Morbidelli, A., Rickman, H.\ 
Comets as collisional fragments of a primordial planetesimal disk.\ {\it Astron. Astrophys.} 
{\bf 583}, A43 (2015).

\bibitem[Wetherill(1967)]{1967JGR....72.2429W} [43] Wetherill, G.~W.\ Collisions in the 
asteroid belt.\ {\it J. of Geophys. Research} {\bf 72}, 2429 (1967).

\bibitem[Greenberg(1982)]{1982AJ.....87..184G} [44] Greenberg, R.\ Orbital interactions - 
A new geometrical formalism. {\it Astron. J.} {\bf 87}, 184-195 (1982).

\bibitem[Davis et al.(2002)]{2002aste.book..545D} [45] Davis, D.~R., Durda, D.~D., Marzari, F., 
Campo Bagatin, A., Gil-Hutton, R.\ Collisional Evolution of Small-Body Populations.\ 
{\it Asteroids III}, 545-558 (2002).

\bibitem[Dell'Oro and Cellino(2007)]{2007MNRAS.380..399D} [46] Dell'Oro, A., Cellino, A.\ The 
random walk of Main Belt asteroids: orbital mobility by non-destructive collisions.\ 
{\it Mon. Not. R. Astron. Soc.} {\bf 380}, 399-416 (2007).

\end{thebibliography}
\end{document}